\documentclass{osa-article}

\journal{osajournal}

\articletype{Research Article}

\begin{document}

\title{Silicon rich nitride-lithium niobate on insulator platform for photonics integrated circuits}

\author{Yang Liu,\authormark{1,2,$\dagger$} Xingrui Huang,\authormark{1,2,$\dagger$} Zezheng Li,\authormark{1,2} Huan Guan,\authormark{1} Zhongchao Fan,\authormark{3,4} Weihua Han,\authormark{3,4} and Zhiyong Li\authormark{1,*}}

\address{\authormark{1}State Key Laboratory on Integrated Optoelectronics, Institute of Semiconductors, Chinese Academy of Sciences, Beijing 100083, China\\
\authormark{2}College of Materials Science and Opto-Electronic Technology, University of Chinese Academy of Sciences, Beijing 100083, China\\
\authormark{3}School of Electronic, Electrical and Communication Engineering, University of Chinese Academy of Sciences, Beijing 101408, China\\
\authormark{4}Engineering Research Center for Semiconductor Integrated Technology, Institute of Semiconductors, Chinese Academy of Sciences, Beijing 100083, China}
\email{\authormark{*}lizhy@semi.ac.cn} 



\begin{abstract}
In this paper, a silicon rich nitride-lithium niobate on insulator (SRN-LNOI) hybrid platform with waveguides and several key components is proposed. The propagation loss of the silicon rich nitride-lithium niobate rib-loaded waveguide (1 $\mu m$ $\times$ 300 nm) is 2.6 dB/cm at 1550 nm. Passive devices, including adiabatic power splitters, multimode interferometer based splitters, asymmetrical Mach-Zehnder interferometers and  Bragg grating filters are fully designed and characterized. Moreover, we report the first demonstration of Mach-Zehnder modulators based on LNOI with high-speed modulation up to 120 GBaud without digital compensation. Hence, the proposed platform enables high performance passive and active devices with low loss, high integration density and complementary metal-oxide-semiconductor technology (CMOS) compatibility, making it a promising candidate for emerging photonics integrated circuits. 
\end{abstract}

\section{Introduction}
In the last decades, the demands for high speed optical communication have led to a significant interest in photonic integrated circuits (PIC). The PIC, which integrates active and passive optical components on a single chip, has been developed on several platforms, including silicon on insulator (SOI) \cite{RN1415}, silicon nitride (SiN) \cite{RN1418}, indium phosphate (InP) \cite{RN1420}, and lithium niobate on insulator (LNOI) \cite{RN1421}. Among those platforms, LNOI is a promising PIC platform because of its strong electro-optical coefficient ($r_{33}$ = 30.8 pm/V at $\lambda$ = 630 nm) and high refractive index contrast ($\Delta n \approx 0.7$ at $\lambda$ = 1550 nm) \cite{RN1424}. Up to now, various high-performance optical devices such as modulators, low-loss waveguides, and micro-ring resonators fabricated by constructing microstructures on the LNOI platform have been reported. And such tremendous progress benefits from the rapid development of manufacturing methods for LNOI microstructures.

Generally, there are two mainstream manufacturing methods for LNOI microstructures: 1) direct etching of LNOI (monolithic integration) and 2) combining the lithium niobate (LN) thin film with other materials (hybrid integration). For the monolithic integration, waveguides with propagation losses of 5 dB/cm  \cite{RN1429}, $3 \pm 0.2$ dB/cm \cite{RN1431}, 0.4 dB/cm \cite{RN1128}, 0.15 dB/cm \cite{Ying:21} and, 2.7 dB/m  \cite{RN1433} are demonstrated in pioneering works. And high performance modulators with 3 dB bandwidth of 100 GHz \cite{RN1208}, 60 GHz \cite{Ying:21} and >100 GHz \cite{RN1509} have been reported. However, because of the chemical stability of LN, the direct etching approach has strict requirements for the laboratory equipment and often results in non-vertical waveguide sidewalls. 

Considering challenges in the direct etching approach, hybrid integration is an attractive choice. On the hybrid LNOI platform, a waveguide is formed by loading a strip of other material on LN thin films as the core, or bonding LN thin films onto other materials as the top cladding  \cite{RN1424}. Several remarkable studies have been made by the bonding approach. For instance, Cai et al. have reported a series of high-performance photonics devices, including a grating coupler with 3.06 dB coupling efficiency  \cite{RN1434}, a Michelson interferometer modulator with 3 dB bandwidth of 17.5 GHz  \cite{RN1437}, and a Mach-Zehnder modulator with 3 dB bandwidth of 70 GHz \cite{RN1438}, by bonding LN films on the SOI platform. But the bonding approach will come up with complex fabrication and reliability problems under high temperature. 

Besides, an alternative is loading LN thin film with a  of material that has a similar refractive index, such as SiN \cite{RN1448}, tantalum pentoxide ($Ta_{2}O_{5}$) \cite{RN1441}, and chalcogenide glass \cite{RN1443}. And a high-efficiency modulator with 3 dB bandwidth of 40 GHz has been proposed on the 
SiN-LNOI platform \cite{RN1508}. However, the refractive indexes of those materials are lower than LN and the mode field mainly distributes in the LN film, which leads to a number of disadvantages, such as large bending radius (> 200 $\mu m$), difficulties in regulating the optical profile and the lack of efficient  input/output components.

To address those issues, we propose an ultra-compact and CMOS-competitive solution: the silicon rich nitride-lithium niobate (SRN-LNOI) platform. In this work, we present the SRN-LNOI platform with rib-loaded waveguides and several basic components for PIC. In Section 2, the design and fabrication of the silicon rich nitride-lithium niobate (SRN-LN) rib-loaded waveguide are presented. In Section 3, compact passive devices, including power splitters, Bragg grating filters, and asymmetric Mach–Zehnder interferometers (AMZI) are proposed. Then, we demonstrated a Mach-Zehnder modulator with a recorded high symbol rate of 120 GBaud in Section 4. And Section 5 presents the conclusion of this research.

\section{SRN-LN rib-loaded waveguide}
\begin{figure}[h!]
\centering
\includegraphics[width=10cm]{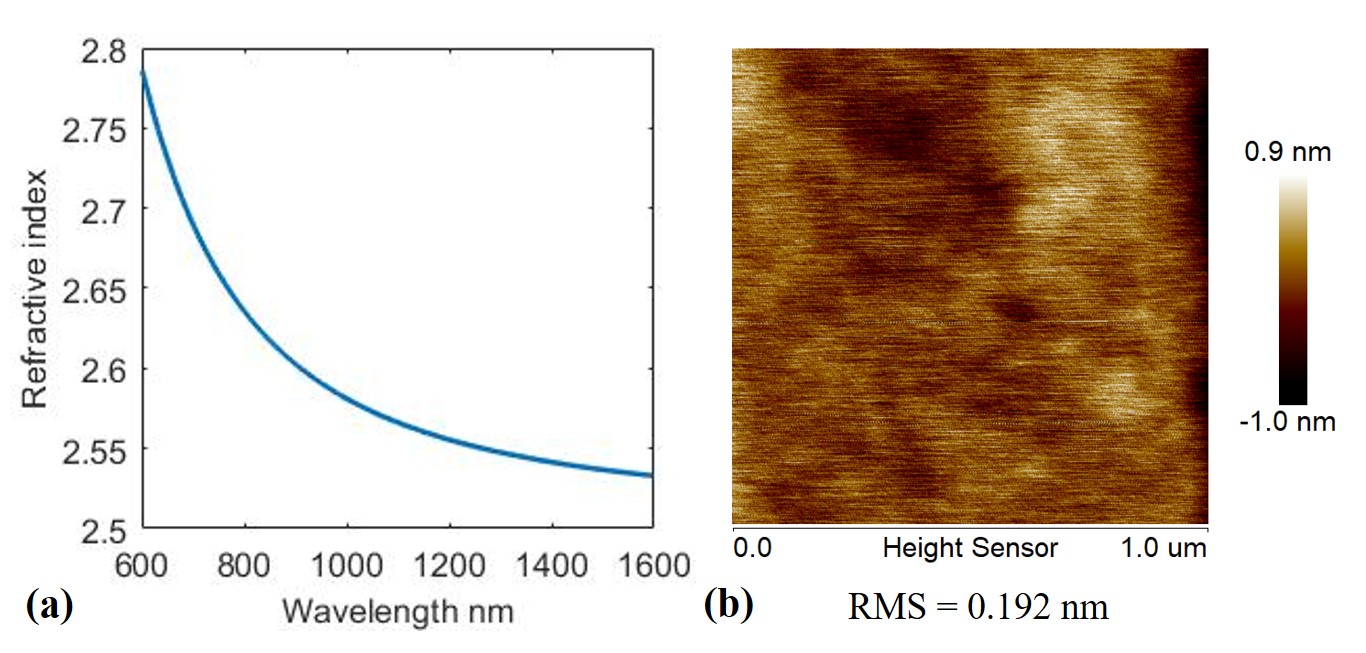}
\caption{(a) Refractive index dispersion
curve of the proposed SRN. (b) AFM micrographs with the RMS roughness of SRN films.}
\label{fig:1}
\end{figure}

The first step to build up the SRN-LN rib-loaded waveguide is the deposition of SRN \cite{RN1445}. An X-cut LNOI wafer with 400 nm thin film LN and 4.7 $\mu m$ buried oxide (BOX) layer (from NanoLN) was used as the substrate. Thin film SRN was deposited by Oxford Plasma Technology 100 plasma enhanced chemical vapor deposition (PECVD) system, employing the deposition recipe as following: $SiH_4$ (4 sccm)/ $N_2$ (800 sccm), chamber pressure of 800 mTorr, RF power of 30 W, and temperature of 300 °C. The recipe was optimized to obtain an appropriate refractive index and a small roughness. It is worth noting that 50 nm silica was deposited between the LN layer and SRN layer to improve the thickness uniformity of SRN films. The ellipsometer measurement presented in Fig. \ref{fig:1} (a) shows that the refractive index of the SRN layer is 2.53 (at 1550 nm). Moreover, other refractive indexes of SRN can be achieved by changing the N/Si ratio of SRN. The root mean square (RMS) roughness, which is one of the main sources of scattering losses, is as low as 0.192 nm measured by the atomic force microscope (AFM). It is the smallest RMS roughness among the reported SRN films to the best of our knowledge. The SRN film with a high refractive index and a small roughness lays down a solid foundation for SRN-LN rib-loaded waveguides.

\begin{figure}[htbp]
\centering
\includegraphics[width=10cm]{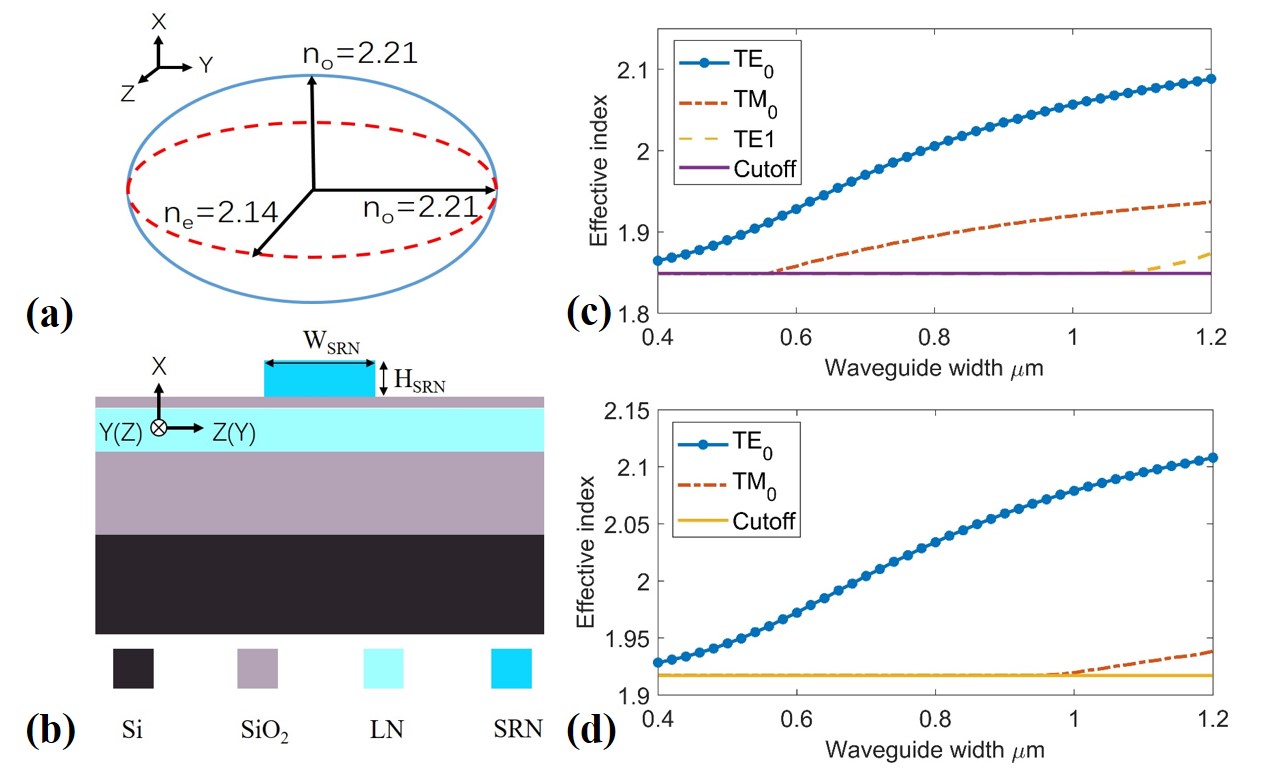}
\caption{(a) The refractive index ellipsoid of LN. (b) Schematic of the proposed SRN-LN rib-loaded waveguide. Effective index with different waveguide widths along crystallographic Y (c) and Z (d) directions. }
\label{fig:2}
\end{figure}

Based on the smooth SRN film with a refractive index of 2.53, a SRN-LN rib-loaded waveguide was designed and fabricated. The waveguide direction plays an important role in the waveguide design because of the inherent birefringence of LN. From the Fig. \ref{fig:2} (a) we can see that the LN exhibits a strong birefringence with an extraordinary refractive index ($n_e$) of 2.14 and ordinary refractive index ($n_o$) of 2.21 at 1550 nm. For example, the refractive index of the LN layer is $n_e$ ($n_o$) with the waveguide direction along the crystallographic Y (Z) direction for the transverse electric (TE) mode. As a result, the waveguides along both of the crystallographic Y and Z directions were considered. The schematic of the SRN-LN rib-loaded waveguide is shown in Fig. \ref{fig:2} (b). The SRN-LN rib-loaded waveguide is composed of a silicon substrate layer, a BOX layer, a LN layer, a silica buffer layer, and a SRN layer from bottom to top. The main parameters of the waveguide are the waveguide width ($W_{SRN}$) and waveguide height ($H_{SRN}$). The $H_{SRN}$ of 300 nm was selected considering the moderate optical confinement. A finite difference eigenmode solver (Lumerical MODE Solutions) was utilized to analyze the single-mode condition of SRN-LN rib-loaded waveguides. Figure \ref{fig:2} (c) and (d) show effective indexes with different $W_{SRN}$ along crystallographic Y and Z directions of LN layer. The effective index rises with the increasing of $W_{SRN}$. And the single-mode condition can be maintained with $W_{SRN}$ < 1.1 $\mu m$ along both crystallographic Y and Z directions. 

\begin{figure}[htbp]
\centering
\includegraphics[width=10cm]{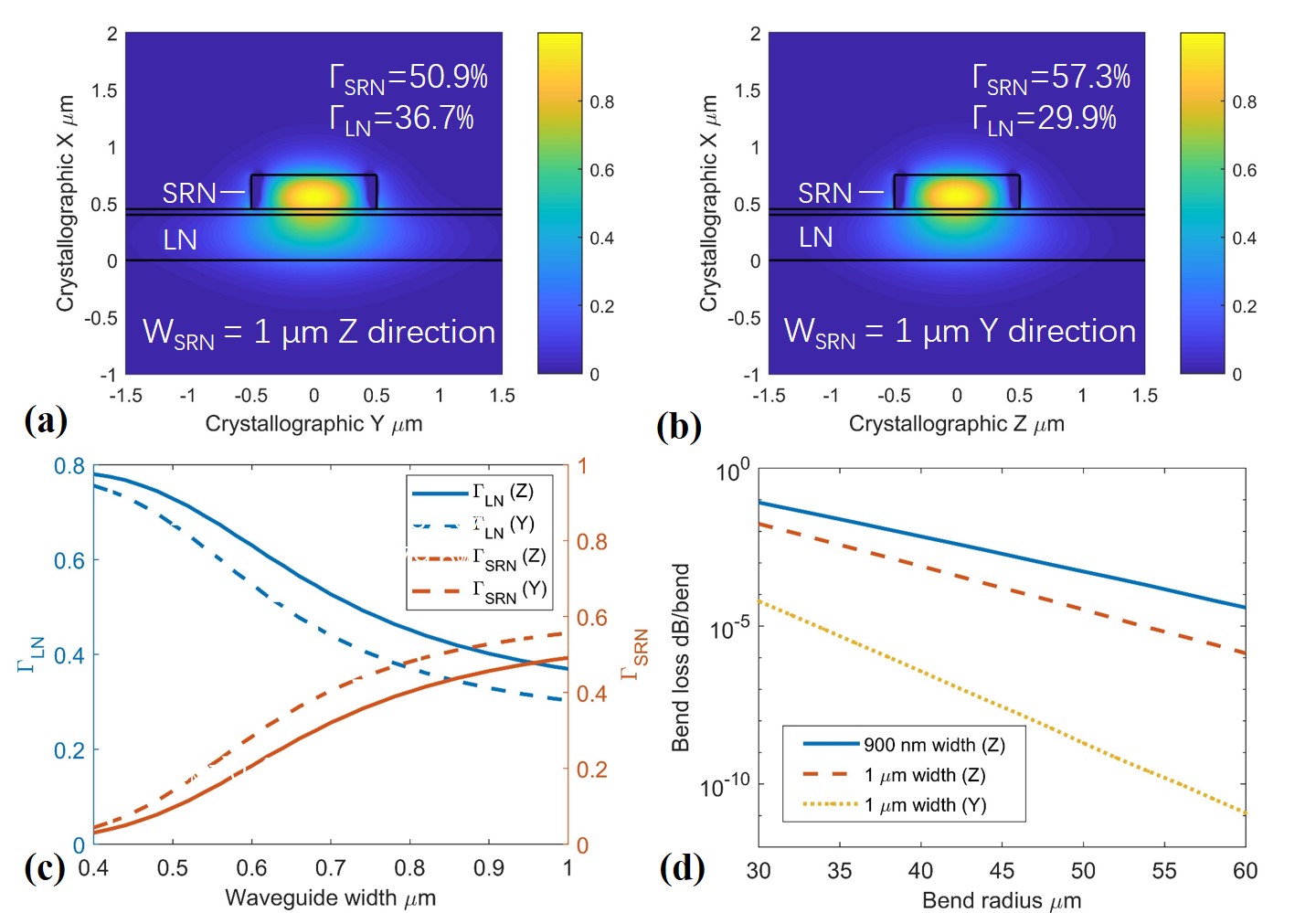}
\caption{Mode field distributions of waveguides with width of 1 $\mu m$ along crystallographic Z (a) and Y (b) direction. (c) Mode field distributions of waveguides with width of 0.4 $\mu m$ along crystallographic Z direction. (d) Simulated bend loss versus the radius.}
\label{fig:2-1}
\end{figure}

Despite the single-mode condition, the $W_{SRN}$ and the waveguide direction impact greatly on the optical confinement and the minimum bend radius of waveguides as well. In order to identify the influences of $W_{SRN}$ clearly, a comparison of waveguides with different $W_{SRN}$ were performed. Firstly, the birefringence was taken into account. Figure \ref{fig:2-1} (a) and (b) depicte the simulated mode fileds of waveguides with $W_{SRN}$ = 1 $\mu m$ along the crystallographic Z and Y direction. The mode fileds are mostly confined in the SRN strip. The confinement fraction in the LN layer ($\Gamma_{LN}$) and the SRN layer ($\Gamma_{SRN}$) are 36.7\% (29.9\%) and 50.9\% (57.3\%) along the crystallographic Z (Y) direction. Patently, the waveguides (Y direction) have better optical confinement than the waveguides (Z direction). The influences of $W_{SRN}$ were then analyzed. The Fig. \ref{fig:2-1} (c) presents the $\Gamma_{LN}$ and $\Gamma_{SRN}$ of waveguides with different $W_{SRN}$ along the crystallographic Z and Y directions. As the $W_{SRN}$ increasing, the $\Gamma_{LN}$ will decrease and the $\Gamma_{SRN}$ will rise. This result indicates that the optical confinement decreases with the decline of $W_{SRN}$. And the weaker optical confinement is, the larger the bend radius of the waveguide will be. Figure \ref{fig:2-1} (d) shows the bend loss of waveguides with different widths and directions. For the crystallographic Z direction, the bend losses of waveguides with $W_{SRN}$ = 1 and 0.9 $\mu m$ sharply drop with the increase of bend radius. And a wider $W_{SRN}$ results in a smaller bend radius. Meanwhile, the bend loss of waveguides (Y direction) is smaller than waveguides (Z direction) under the same structures. Overall, waveguides with radius of 50 $\mu m$ and $W_{SRN}$ of 1 $\mu m$ can achieve a negligible bend loss in both crystallographic Y and Z directions. Consequently, $W_{SRN}$ of 1 $\mu m$ was chosen to maintain low waveguide loss and single-mode condition.

\begin{figure}[htbp]
\centering
\includegraphics[width=10cm]{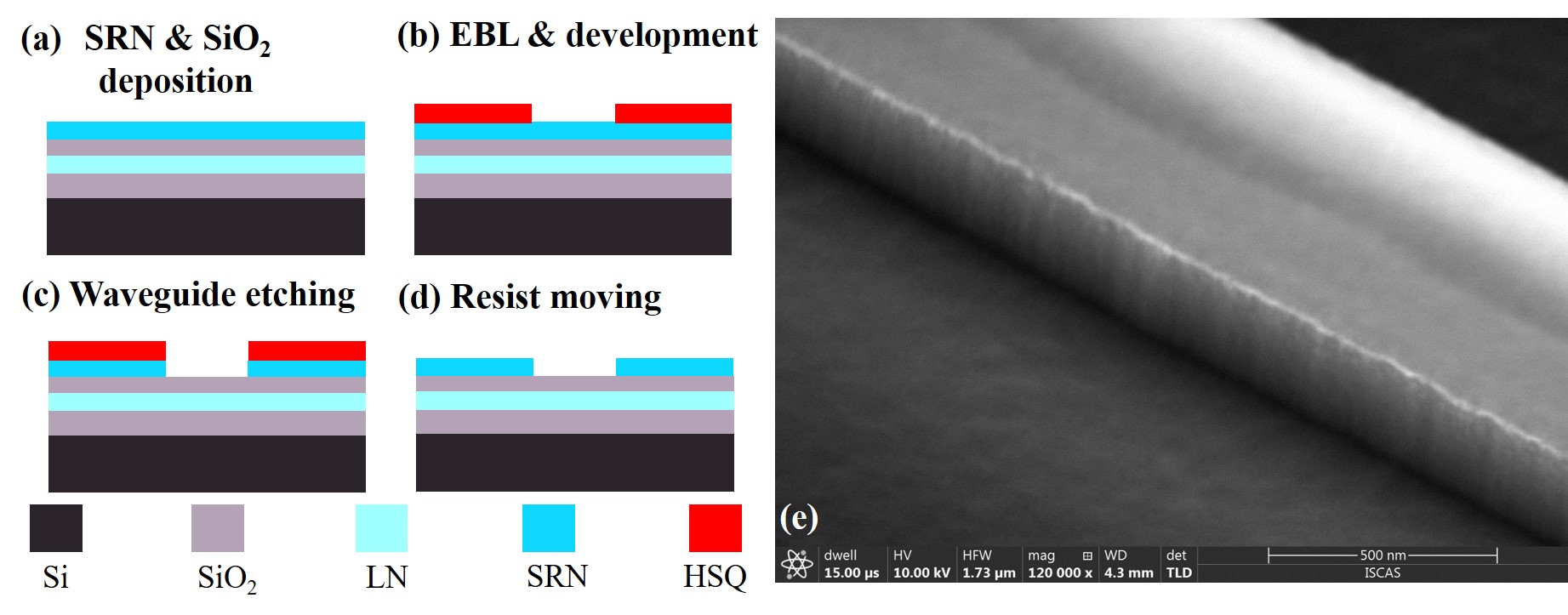}
\caption{(a)-(d) Fabrication process of SRN-LN rib-loaded waveguides. (e) SEM image of the fabricated SRN-LN rib-loaded waveguides.}
\label{fig:3}
\end{figure}

Figure \ref{fig:3} (a)-(d) shows the fabrication process of the SRN-LN rib-loaded waveguide. To define the waveguides, electron beam photoresist hydrogen silsesquioxane (HSQ, XR1541) was spin-coated on the SRN layer and soft baked on an 80 °C hot plate for 4 min. Afterward, HSQ was patterned by electron beam lithography (EBL, Raith e-Line plus). The patterns were then transferred to the SRN film using fluorine-based reactive ion etching (RIE). Finally, the fabrication of the SRN-LN rib-loaded waveguides were completed after removing the resist, and the scanning electron microscope (SEM) image of the fabricated waveguide is shown in Figs. \ref{fig:3} (e).

\begin{figure}[htbp]
\centering
\includegraphics[width=10cm]{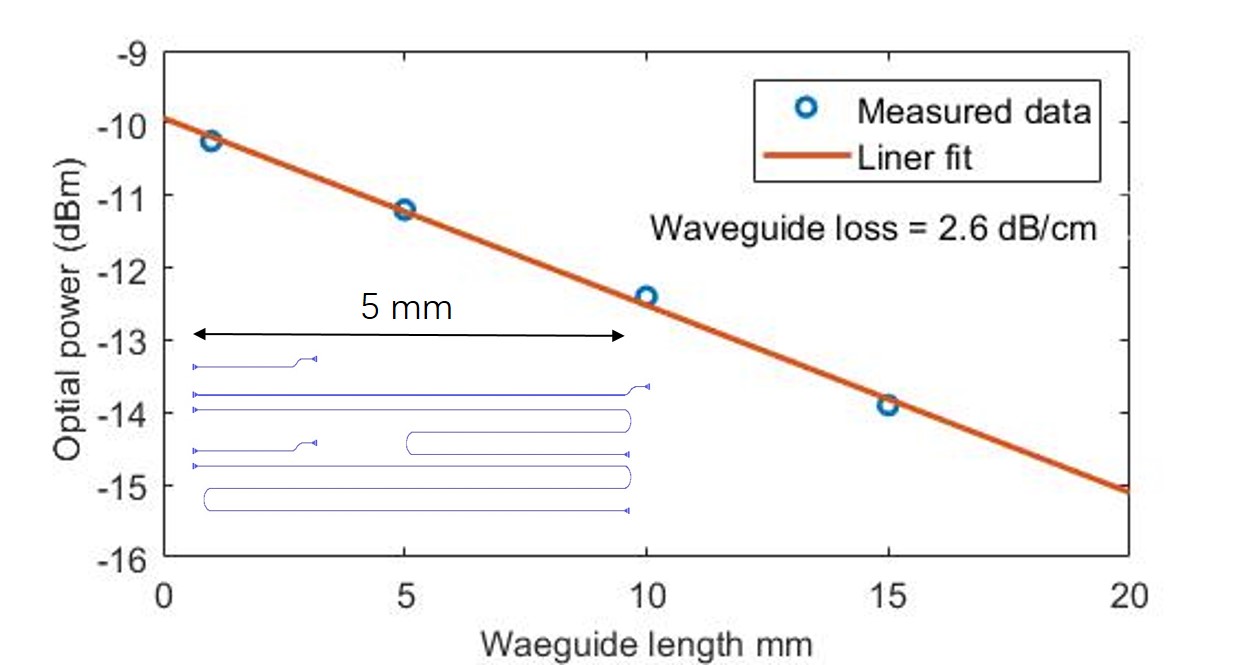}
\caption{Measured propagation loss and linear fitting line. Insertion image: layout of the meandered waveguides.}
\label{fig:4}
\end{figure}

The propagation loss of the SRN-LN rib-loaded waveguide was measured by the cut-back method. Waveguides with different lengths (1-15 mm) were fabricated and measured. A tunable laser working at 1550 nm was used as the source. The light was coupled from fiber to waveguides by utilizing the grating couplers based on our previous work \cite{RN1446}, and the optical output was measured by an optical spectrum analyzer. For 1-$\mu m$ wide waveguides, the propagation loss of the fundamental TE mode is 2.6 dB/cm, as presented in Fig. \ref{fig:4}. As far as we know, the propagation loss of SRN waveguides is comparable to the lowest loss of PECVD SiN waveguides proposed in Ref. \cite{RN1450,RN1451}. Consequently, SRN-LN rib-loaded waveguides can provide low-loss optical interconnection as well as high integration density for various functional devices.

\section{Passive devices}
In this section, several functional passive components for PIC, including optical power splitters, and filters are designed, fabricated, and tested. The finite difference time domain method (FDTD, Lumerical) is used to simulate the passive devices. Besides, the fabrication process and measurement approach are the same as those of the SRN-LN rib-loaded waveguides mentioned above. The light is coupled in and out from the passive devices using the grating couplers based on our previous work \cite{RN1446}.

\subsection{Power splitters}
\begin{figure}[h!]
\centering
\includegraphics[width=8cm]{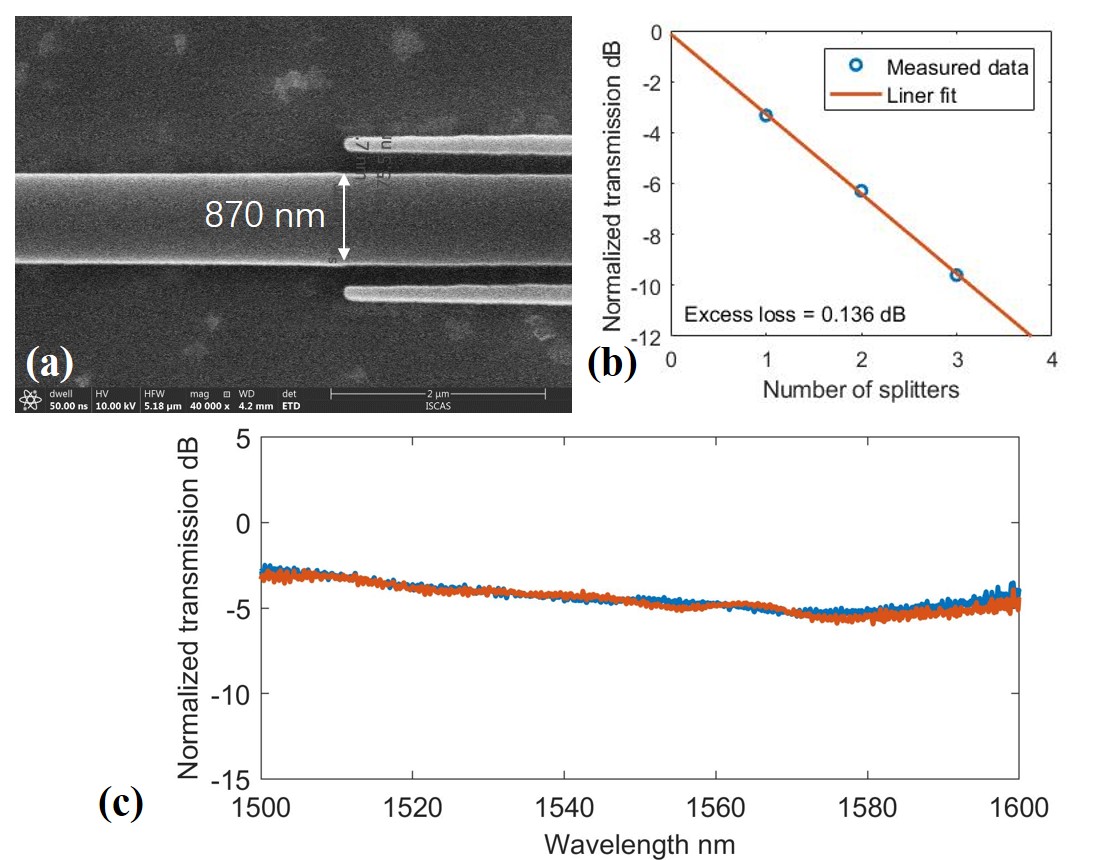}
\caption{(a) SEM image of the fabricated adiabatic power splitter. (b) Measured excess loss of adiabatic power splitters. (c) Measured spectrum of the adiabatic power splitter.}
\label{fig:5}
\end{figure}

Optical power splitters (OPS) are essential components for optical power distributions in PIC. Various power splitters of different structures have been realized, including Y-branches \cite{RN1295}, directional couplers (DC) \cite{RN1296}, bent directional couplers (bent DC) \cite{RN1297}, multimode interferometers (MMI) \cite{RN1301}, and adiabatic taper \cite{RN1294}. The Y-branches power splitters are the simplest structures but introduce large mode mismatch loss or large device length. The DC and bent DC structures require for long device length (50 $\mu m$). Comparing with those structures, the adiabatic power splitter with adiabatic taper is an attractive option for its low excess loss (EL), small footprint, and broad bandwidth. 

We experimentally demonstrate the adiabatic power splitter (APS) on the SRN-LNOI platform. Figure \ref{fig:5} (a) shows the structure of the APS, which consists of one input taper and two symmetrical output tapers. The parameters of the APS are the input waveguide width of 900 nm, the tip width of 150 nm, the taper length of 20 $\mu m$, and the gap of 200 nm between two adjacent tapers. After fabrication, an adiabatic power splitter with the input waveguide width of 870 nm, the tip width of 115 nm, the gap of 175 nm, and the taper length of 20 $\mu m$ was obtained. To determine the EL of APS, three cascaded splitters were measured, and the results are presented in Figure \ref{fig:5} (b). By linear fitting, the EL of The APS is about 0.136 dB. The optical spectrum of the single APS indicate that uniform outputs are achieved from 1500 nm to 1600 nm, as shown in Fig. \ref{fig:5} (c).

\begin{figure}[h!]
\centering
\includegraphics[width=8cm]{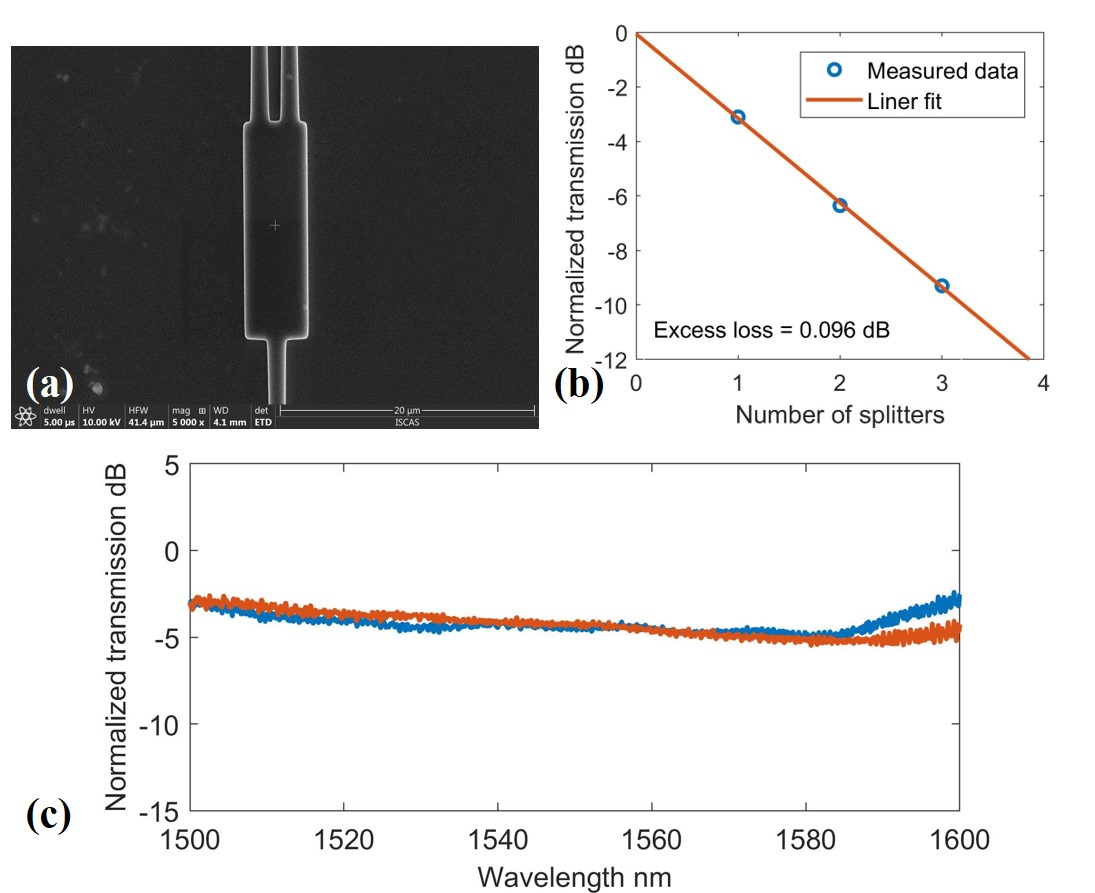}
\caption{(a) SEM image of the MMI based power splitter. (b) Measured excess loss of MMI based power splitters. (c) Measured spectrum of the MMI based power splitter.}
\label{fig:6}
\end{figure}

Moreover, a multimode interferometer (MMI) based optical power splitter is demonstrated. The SEM image of MMI splitter with a multimode waveguide length of 16.8 $\mu m$ and width of 4.8 $\mu m$ is shown in Fig. \ref{fig:6} (a). To reduce the EL, the input and output waveguides are linearly tapered from 1 $\mu m$ to 1.5 $\mu m$ before connecting to the multimode waveguide. The measurement method is the same as the APS. As shown in Fig. \ref{fig:6} (b), the EL of 0.096 dB is obtained. Figure \ref{fig:6} (c) presents the optical spectrum of an MMI splitter. Uniform outputs are achieved from 1500 nm to 1580 nm.

\subsection{filters}
\begin{figure}[htbp]
\centering
\includegraphics[width=10cm]{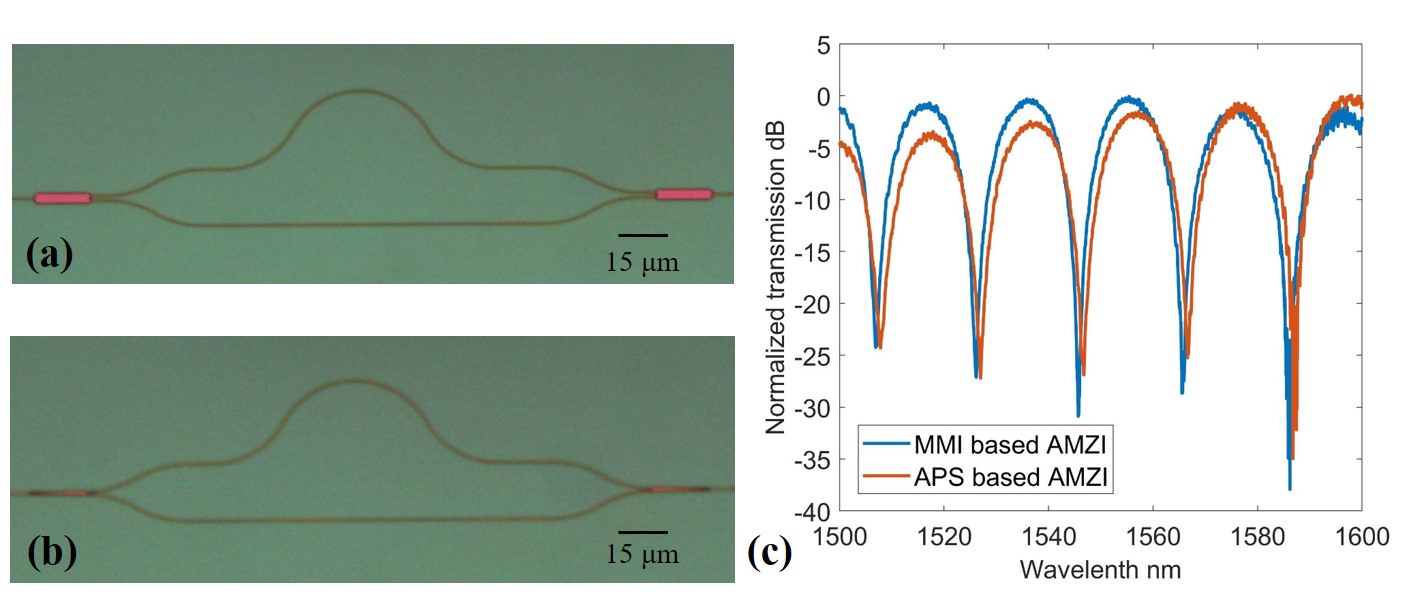}
\caption{Microscope images of AMZIs based on the MMI splitter (a) and the APS (b). (c) Measured spectrum of the AMZIs.}
\label{fig:7}
\end{figure}

Apart from optical splitters, filters play an important role in PIC. Based on the proposed optical power splitters, AMZIs were fabricated with arm length differences of 50 $\mu m$. The microscope image of the proposed AMZIs based on MMI splitter and APS are shown in Fig. \ref{fig:7} (a) and (b), respectively. Figure \ref{fig:7} (c) shows the transmission spectrum of the AMZI and the extinction ratio (ER) of both AMZIs are larger than 20 dB from 1500 nm to 1600 nm. The high extinction ratio also proves that the high power uniformity of the proposed optical power splitter is achieved.

\begin{figure}[htbp]
\centering
\includegraphics[width=10cm]{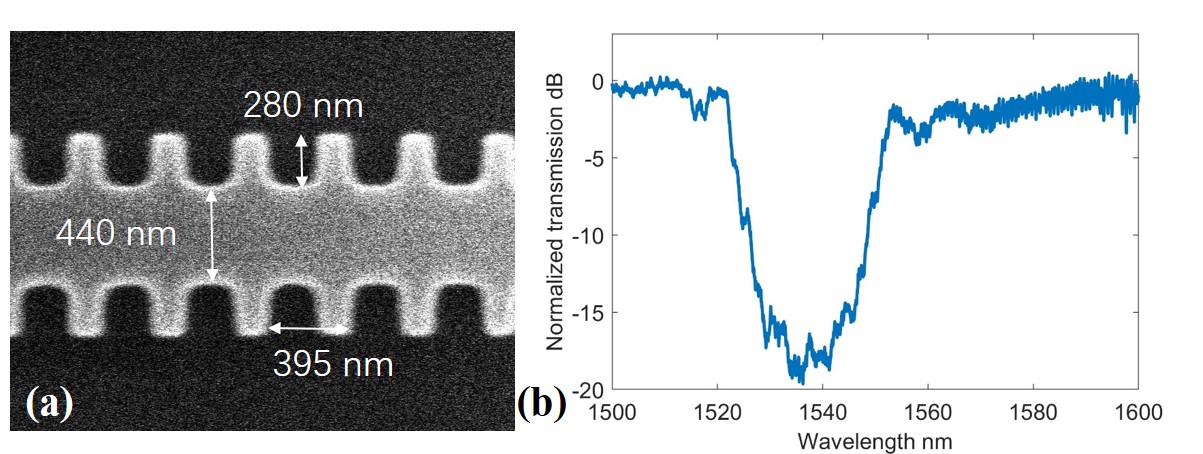}
\caption{(a) SEM image of the fabricated Bragg grating filter. (b) Measured spectrum of the Bragg grating filter.}
\label{fig:8}
\end{figure}

Meanwhile, a Bragg grating filter was also designed and fabricated. The Bragg grating filter has a width of 1 $\mu m$, a grating period of 395 nm, a period number of 100, a grating length of 39.5 $\mu m$, a side corrugation depth of 250 nm, and a teeth width of 185 nm. Figure \ref{fig:8} (a) shows the SEM image of the Bragg grating filter. Because of the fabrication error, a Bragg grating filter with a side corrugation depth of 280 nm, and a teeth width of 160 nm was obtained. The deviation of 25-30 nm from the designed value leads to a blue shift of the center wavelength compared with the simulation. Figure \ref{fig:8} (b) shows the measured transmission of the Bragg grating filter. And a Bragg grating filter with center wavelength of 1538 nm, ER of 19 dB, and 3 dB bandwidth of 14 nm was achieved with a period number of 100. By contrast, a Bragg grating filter based on monolithic LNOI platform exhibits a ER of 14 dB with a period number of 1000 \cite{RN1452}. This result suggests that the SRN-LNOI platform is more suitable for high perfoemance Bragg grating filters with high ER and compact footprints.

\section{Mach–Zehnder modulators}
\begin{figure}[htbp]
\centering
\includegraphics[width=10cm]{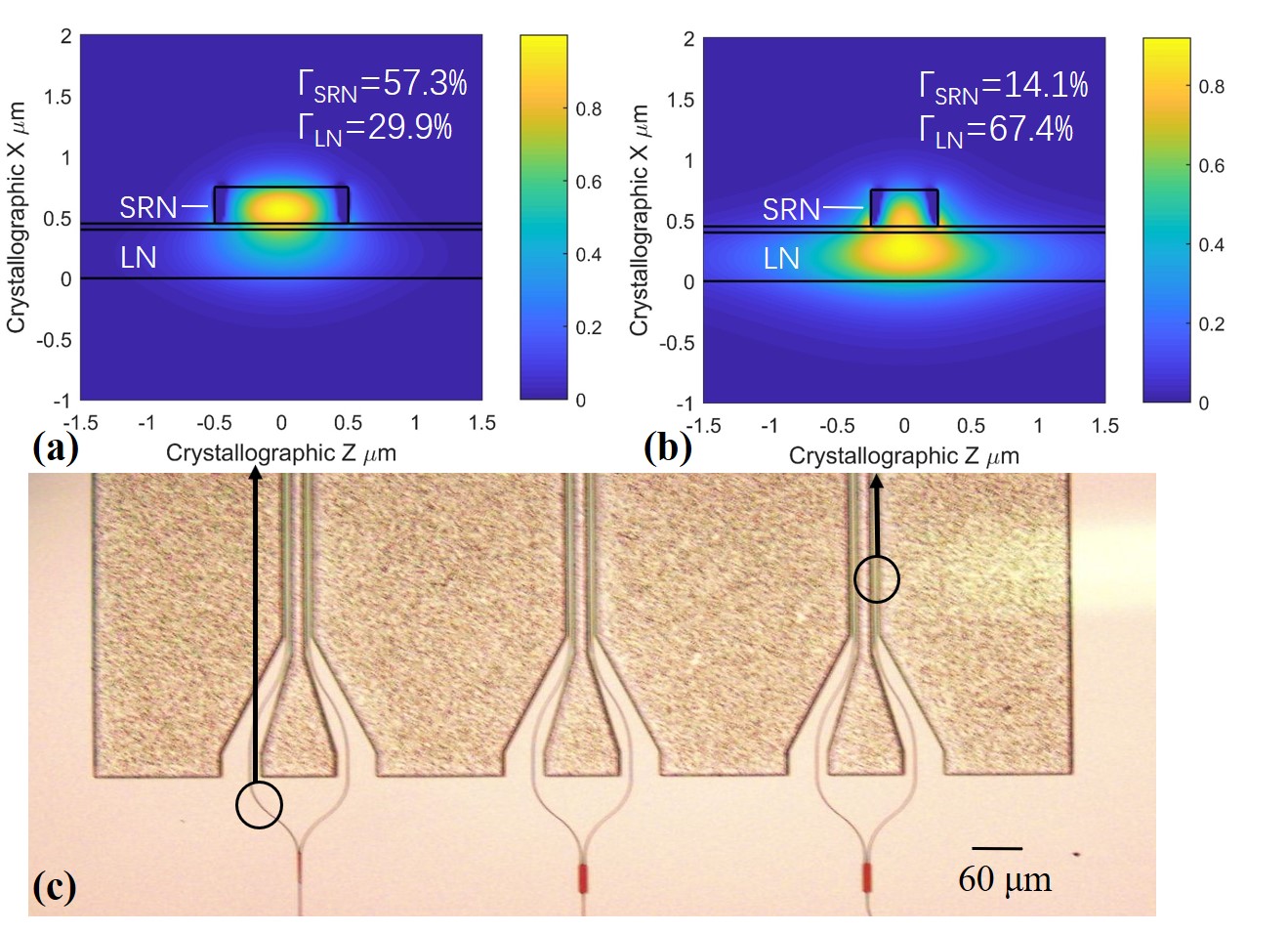}
\caption{Mode field distributions of waveguides with width of 1 $\mu m$ (a) and 500 nm (b). (c) Microscope images of the proposed Mach–Zehnder modulators.}
\label{fig:9}
\end{figure}

LN is a promising material for electro-optical modulation because of its strong electro-optical coefficient ($r_{33}$ = 30.8 pm/V at $\lambda$ = 630 nm). In this section, a high performance Mach–Zehnder modulator (MZM) is demonstrated on the SRN-LNOI hybrid platform. As mentioned above, an x-cut LNOI wafer with 400 nm thin-film LN and 4.7 $\mu m$ BOX layer wafer was used as the substrate, and 300 nm SRN was deposited as the waveguide layer. The optical part of MZM comprises two regions: the passive region and the active region. The passive region consists of waveguides (1 $\mu m$ $\times$ 300 nm), bend waveguides with 50 $\mu m$ radius and optical splitters proposed in section 3.1. At the passive region, the optical power is mostly confined in the upper SRN waveguide. As shown in Fig \ref{fig:9}. (a), the confinement fraction in the LN layer ($\Gamma_{LN}$) and the SRN layer ($\Gamma_{SRN}$) are 29.9\% and 57.3\%, respectively. At the active region, a 5 mm long waveguides with width of 500 nm and height of 300 nm are applied. As Fig. \ref{fig:9} (b) shows, light is almost guided in the LN slab layer, with $\Gamma_{LN} = 67.4\%$, $\Gamma_{SRN} = 14.1\%$, and group index ($n_g$) of 2.373. As a result, the large $\Gamma_{LN}$ enables efficient modulation at the active region. A 100 $\mu m$ long adiabatic waveguide taper is utilized to connect the passive and active regions.

In the active region of the modulator, a ground-signal-ground (GSG) traveling wave electrode is utilized to perform the single-drive push-pull operation, as shown in Fig. \ref{fig:9} (c). After sputtering a 30 nm chromium adhesion layer, 1 $\mu m$ aluminum was sputtered and patterned by photolithography. The signal electrode width is 16 $\mu m$, and the ground electrodes widths are 200 $\mu m$ with the electrode gap of 6 $\mu m$. The gap of 6 $\mu m$ is optimized to balance the optical loss caused by aluminum electrodes and modulation efficiency. The electrode transmission line characteristic impedance ($Z_{0}$) and microwave index ($n_{m}$) are 50 $\Omega$ and 2.04, respectively. As we know, the maximal 3-dB electrical roll-off frequency ($f_{3dB, el}$) can be obtained by  matching the velocity between the optical wave and microwave, matching the characteristic impedance to the load and source impedance, and minimizing the RF loss. Assuming impedance matching and no RF loss \cite{RN1447}, the theoretical bandwidth limitation is given as: $f_{3dB}$  (GHz) = $0.13/(L|n_m-n_g|)$, where L is the length of phase-shifter arm. Consequently, the bandwidth limitation of the proposed MZM is 79 GHz with L = 5 mm and $(|n_m-n_g|)$ = 0.33. Moreover, it is worthy to note that the bandwidth limitation caused by velocity mismatch can be optimized by changing the LNOI wafer with a thinner BOX layer.

\begin{figure}[htbp]
\centering
\includegraphics[width=12cm]{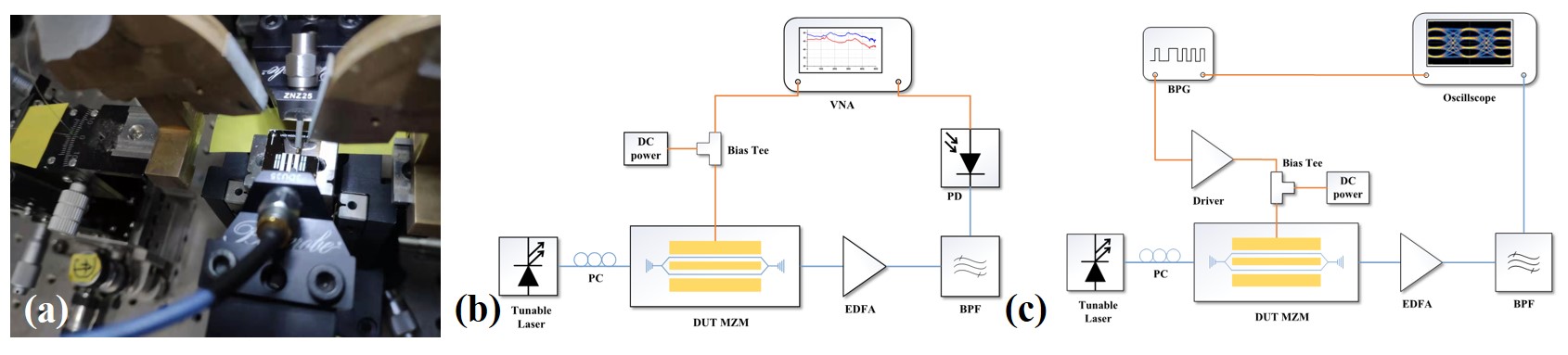}
\caption{(a) Testing image of the Mach–Zehnder modulators. (b) Experimental setup for small signal measurements. (c) Experimental setup for large signal measurements.}
\label{fig:10}
\end{figure}

\begin{figure}[htbp]
\centering
\includegraphics[width=12cm]{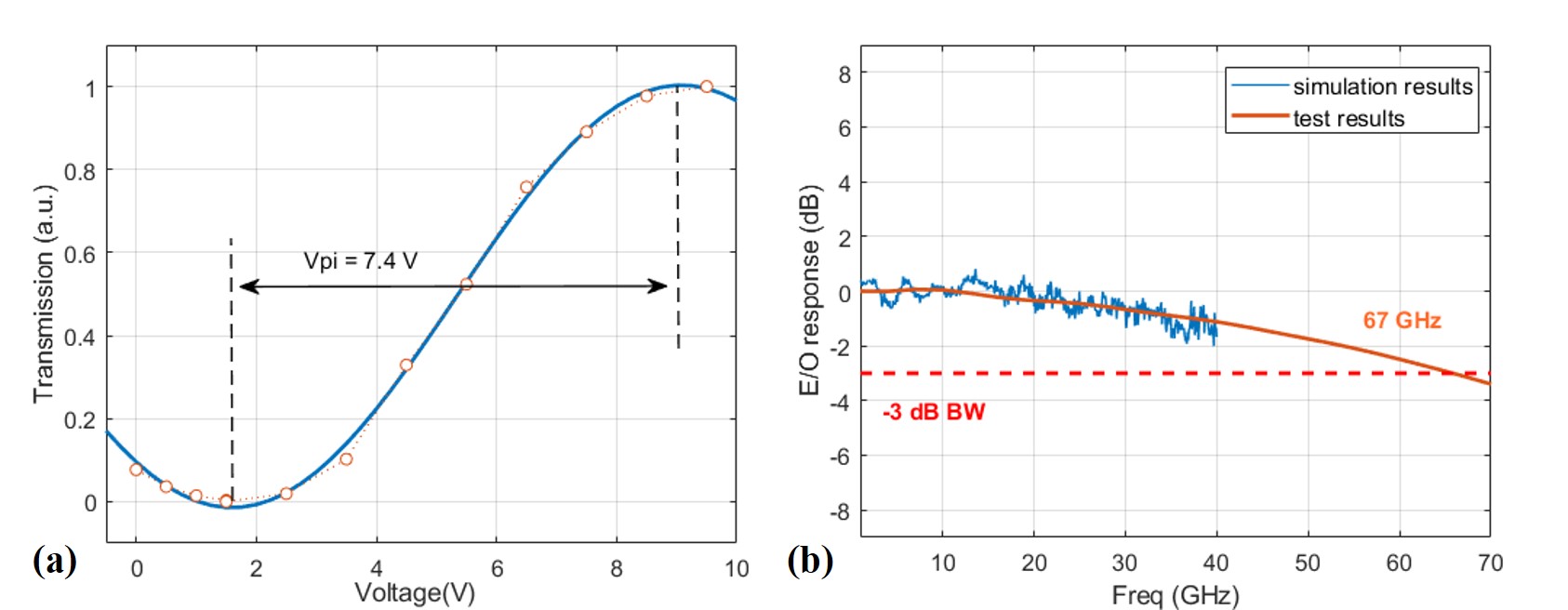}
\caption{(a) Normalized optical transmission as a function of applied voltage on travelling-wave electrodes at 1550 nm. (b) Measured EO S21 parameters under bias voltage of 7 V.}
\label{fig:11}
\end{figure}

After fabricating the MZM, RF measurements including small signal and larger signal characterization were performed, as shown in Figs. \ref{fig:10} (a). Firstly, the DC characterization was measured. As shown in Fig. \ref{fig:11} (a), a ER of 25.8 dB and a modulation efficiency figure-of-merit ($V_{\pi}L$) of 3.7 V.cm were obtained for a 5 mm long device.

Afterward, the characterizations of small signal were performed using a 40 GHz vector network analyzer (VNA, Anritsu MS4640). The light was coupled in and out from the MZM by two grating couplers. Then the optical signal was received by a 50 GHz electro-optical (EO) bandwidth photodetector (PD, Finisar XPDV2150), which was connected with the VNA. A pair of 67 GHz GSG high-frequency RF probes were used to apply RF signal and connect with a 50 $\Omega$ terminator, respectively. By de-embedding the known EO S21 parameters of the PD, the EO bandwidth of the proposed MZM was obtained. The measured EO S21 parameters at the DC bias voltage of 7 V are shown in Fig. \ref{fig:11} (b). The EO response decays about 1.1 dB from DC to 40 GHz and the EO 3 dB-bandwidth is predicted to be 67 GHz.

\begin{figure}[htbp]
\centering
\includegraphics[width=12cm]{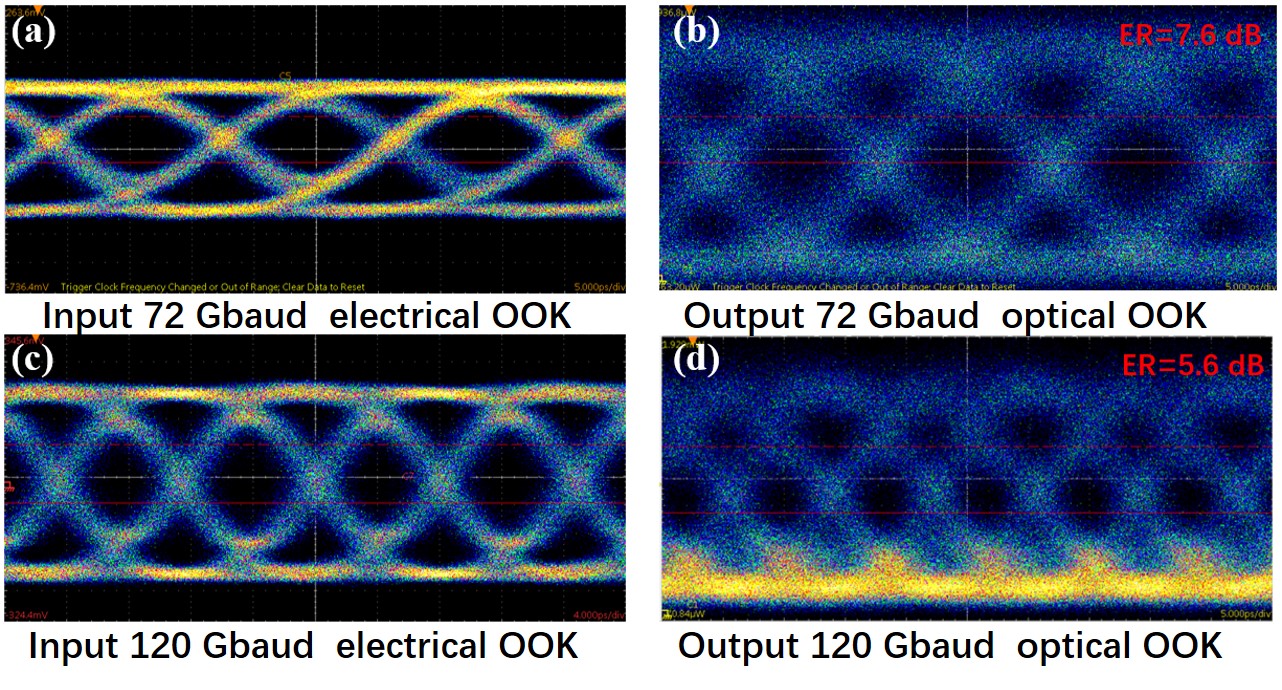}
\caption{Electrical (a) and optical (b) eye diagrams at the rates of 72 GBaud OOK. Electrical (c) and optical (d) eye diagrams at the rates of 120 GBaud OOK.  }
\label{fig:12}
\end{figure}

Then the large signal experiments were conducted as shown in Fig. \ref{fig:10} (c). Firstly, a 72 GBaud RF signal from a bit pattern generator (BPG, SHF 12103A) is applied on the MZM through a 67 GHz GSG microwave probe after amplification. The MZM was set at the quadrature point through the phase shifter. Meanwhile, the modulated optical signal was amplified by an erbium-doped fiber amplifier (EDFA) and filtered using a band-pass filter (BPF). A Tektronix DSA 8300 Digital Serial Analyzer (DSA) with 80 GHz Sampling Module was used to detect the optical signal. Figure \ref{fig:12} (a) and (b) compare the input electrical eye diagram and output optical eye diagram at 72 GBaud on-off keying (OOK) modulation format. Moreover, The BPG and an SHF 2:1 multiplexer (MUX) were used to generate the electrical signals up to 120 GBaud. The input electrical eye diagram and output optical eye diagram at 120 GBaud OOK modulation format are demonstrated in Fig. \ref{fig:12} (c) and (d). The measured ER are 7.6 dB and 5.6 dB for 72 GBaud and 120 GBaud with a 7 V peak-to-peak input voltage, respectively. As far as we know, it is the highest symbol rate achieved in the LNOI MZM, experimentally.

\section{Conclusion}
To draw a conclusion, we present a SRN-LNOI hybrid platform with waveguides and several key components for PIC. Several fundamental components, such as waveguides, optical splitters, and filters were designed, fabricated and tested. A high performance MZM with A EO 3 dB bandwidth of 67 GHz and a recorded high symbol rate of 120 GBaud was experimentally demonstrated. With those proposed fundamental components, the SRN-LNOI hybrid platform provides a solution with high integration density and CMOS compatibility for PIC. 

\begin{backmatter}
\bmsection{Funding}
National Key Research and Development Program of China (Grant No.2018YFE0201900).

\bmsection{Disclosures}
\noindent The authors declare no conflicts of interest.

\medskip

\noindent $^{\dagger}$ These authors contributed equally to this work.
\end{backmatter}


\bibliography{sample}






\end{document}